\documentclass[a4paper,english,a4paper,twocolumn]{revtex4}
\usepackage[T1]{fontenc}
\usepackage[latin9]{inputenc}
\usepackage{graphicx}
\usepackage{amssymb}

\makeatletter

\usepackage{nicefrac}

\makeatletter

\usepackage{graphics}

\usepackage{times}

\usepackage{amstext}

\usepackage{bbm}

\makeatother

\usepackage{babel}
\makeatother

\begin{document}

\title{Coupling strength estimation for spin chains despite restricted access}

\author{Daniel Burgarth$^{1,2}$}

\author{Koji Maruyama$^{2}$}

\author{Franco Nori$^{2,3}$ }

\affiliation{$^{1}$Mathematical Institute, University of Oxford, 24-29 St Giles',
Oxford OX1 3LB, United Kingdom \\
 $^{2}$Advanced Science Institute, The Institute of Physical and
Chemical Research (RIKEN), Wako-shi, Saitama 351-0198, Japan\\
$^{3}$Center for Theoretical Physics, Physics Department, The
University of Michigan, Ann Arbor, Michigan 48109-1040, USA }

\begin{abstract}
Quantum control requires full knowledge of the system many-body Hamiltonian.
In many cases this information is not directly available due to restricted
access to the system. Here we show how to indirectly estimate all the coupling strengths in
a spin chain by measuring one spin at
the end of the chain. We also discuss the efficiency of this `quantum inverse problem` 
and give a numerical example.
\end{abstract}


\maketitle

\section{Introduction}

The great progress in experimental techniques for manipulating microscopic
objects has brought even quantum mechanical systems under control.
It has been one of the strong driving forces to promote the recent
intensive theoretical study of quantum information science. What we
need towards the realization of quantum information processing and
quantum simulations is control over a quantum system that resides
in a high-dimensional Hilbert space. Then, in order to achieve the
physical controllability, we typically decompose the Hilbert space
into a tensor product of many low (usually two-) dimensional spaces
and consider addressing each of them individually as well as coupling
any two members.

Although the addressability of the whole system is thus the key, paradoxically
it is the origin of the biggest problem in quantum control. That is,
that we can access the system means that the surrounding environment
can also interact with it, causing unwanted errors. In order
to circumvent or minimize the effect of noise, various methods have
been proposed, e.g., quantum error correcting codes, decoherence-free
subspaces, and topological quantum computing. Here, we focus on an
alternative approach, i.e., operating only a (small) subset of the
system while isolating the rest from its environment. If we have all
information on the Hamiltonian describing the system, its controllability
can be analyzed by the theory of quantum control (see \cite{D'Alessandro2008}
for an introduction). Recently, it has also been shown that entire
spin chains can be fully controlled by operating on one end only,
provided all parameters of the Hamiltonian are known \cite{Schirmer,Burgarth2008}.

However, if we could completely isolate part of the network from
any interactions with the outer world, then this isolation would prevent
us from obtaining the required information on the shielded subsystem.
While the type of interaction that governs the dynamics inside the
network would be known due to its intended design, the precise values
of various parameters that characterize the full Hamiltonian might
only be in a certain range and not accurate enough to enable us to
have a desired controllability. Hence, a natural question arrises: is it
possible to estimate the necessary parameters of the Hamiltonian only
by accessing a subset of the whole network? We answer this question
positively and show how to do it for chains of spin-1/2 particles
whose interaction is of Heisenberg type. The main task is thus to
estimate the coupling strengths between `untouchable' spins. This
is also useful for quantum state transfer in spin chains, where some
schemes require a good knowledge of the system Hamiltonian \cite{Haselgrove}.

The question of system identification of a quantum device by Fourier
analyis has recently been studied in \cite{Schirmer2}, where the identification
of two-level subspaces was emphasized. Here we give
a solution to system identification for the N-level case under strong
local constraints.
Our work can be seen as an example of inverse problems that are very
important in a number of areas of science and engineering. There are
plethora of situations where indirect probing is the only way to acquire
desired information, from medical ultrasonography to seismic reflection
in geophysics. A closely related problem to the one in this paper
is the estimation of spring constants in classical harmonic oscillator
chains \cite{Gladwell2004}. Our contribution is to provide the quantum
mechanical counterpart of this problem and to discuss the efficiency
of the estimation procedure. Roughly speaking, we obtain information
on the couplings by inserting one excitation into the chain and observing
its return probability. Since the excitation travels back and forth
the chain, it obtains knowledge on \emph{all} the couplings. This
is proved by the main theorem in Section \ref{sec:Setup-and-Main}.
It turns out that in order to obtain good knowledge on the coupling
strengths, the excitation needs to `see' each single link of the chain
$2N$ times, where $N$ is the length of the chain. This claim will
be made rigorous by using the uncertainty principle of the Fourier
transform in Section \ref{sec:Efficiency}.

\begin{figure}
\includegraphics[width=1\columnwidth]{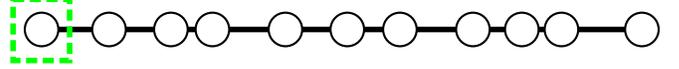}

\caption{(Color online) \emph{\label{fig:2d}All} coupling strengths (black lines) of a chain
of spins can be estimated \emph{indirectly} by quantum state tomography
on one end (dashed area).}

\end{figure}

\section{Setup and Main Result\label{sec:Setup-and-Main}}

We consider a chain of $N$ spins coupled by an anisotropic Heisenberg
Hamiltonian, i.e. \begin{equation}
H=\sum_{n=1}^{N-1}\delta_{n}\left(\sigma_{n}^{+}\sigma_{n+1}^{-}+\sigma_{n}^{-}\sigma_{n+1}^{+}+\Delta\sigma_{n}^{z}\sigma_{n+1}^{z}\right)\label{eq:h2}\end{equation}
with unknown couplings $\delta_{n}$ and anisotropy $\Delta$. Additionally, 
we assume that we know the signs of $\delta_{1}$ and $\delta_{2}.$
$H$ conserves the number of excitations, i.e. \[
\left[\sum_{v\in V}Z_{v},H\right]=0.\]
 This allows us to speak about local excitations in the system. In
particular,  we will see later that for almost all cases it suffices
to restrict the parameter estimation to the \emph{single excitation}
sector of the system. As usual, we let $|\mathbf{n}\rangle$ denote
single excitation states with spin $n$ in the state $|1\rangle$
and all other spins in the state $|0\rangle.$ The state with all
spins in $|0\rangle$ will be denoted as $|\mathbf{0}\rangle.$ The
unknown part of the Hamiltonian is the \emph{interaction strength}
$\delta_{n}$ between spin each neighboring pair of spins $(n,n+1).$
The purpose of the following will be to estimate these coupling strengths.

\begin{description}
\item [{Theorem:}] \emph{Assume that the Hamiltonian of Eq.~(\ref{eq:h2})
has a non-degenerate spectrum in the first excitation sector. Then
the coupling constants $\delta_{n}$ can be obtained by acting on
on the first spin only.} 
\end{description}
We remark that although the non-degenerate spectrum is the generic
case, in practice almost-degenerate eigenvalues can be problematic
to the efficiency of the method suggested below. This will be discussed
in more detail in Section \ref{sec:Efficiency}. The following lemma
is similar to the inverse problems for classical oscillator chains
considered in \cite{Gladwell2004}:

\begin{description}
\item [{Lemma:}] \emph{Assume that all eigenvalues $E_{j}$ $(j=1,\ldots,N$)
in the first excitation sector of the Hamiltonian of Eq. (\ref{eq:h2})
are non-degenerate and known. Assume that for all orthonormal eigenstates
$|E_{j}\rangle$ in the first excitation sector the coefficients $\langle\mathbf{1}|E_{j}\rangle$
are known. Then the coupling constants $\delta_{n}$ are known.}
\end{description}
While the requirements of the lemma may sound unrealistic at the first
sight, we will see towards the end of this section that they are provided
by a simple Fourier analysis of the return probability of a single
excitation.

\paragraph*{Proof of the Lemma:}

Our first observation is that (setting $\delta_{0}=\delta_{N}=0$
for the boundary terms)\begin{equation}
H|\mathbf{n}\rangle=\delta_{n-1}|\mathbf{n-1}\rangle+\delta_{n}|\mathbf{n+1}\rangle+D_{n}|\mathbf{n}\rangle\label{eq:genH}\end{equation}
 with\begin{eqnarray*}
D_{n} & \equiv & \Delta\left(G-2\delta_{n}-2\delta_{n-1}\right)\\
G & \equiv & \sum_{m=1}^{N-1}\delta_{m}.\end{eqnarray*}
The first equation we use is \begin{equation}
D_{n}=\langle\mathbf{n}|H|\mathbf{n}\rangle=\sum E_{j}\left|\langle\mathbf{n}|E_{j}\rangle\right|^{2}.\label{eq:Dn}\end{equation}
In particular this implies that $D_{1}$ is known by the requirements
of the lemma. Then, for all $j$ we have \begin{equation}
E_{j}|E_{j}\rangle=H|E_{j}\rangle.\label{eq:eigen}\end{equation}
Taking the inner product with $\langle\mathbf{n}|$ and using Eq.
(\ref{eq:genH}) we obtain\begin{equation}
\left(E_{j}-D_{n}\right)\langle\mathbf{n}|E_{j}\rangle-\delta_{n-1}\langle\mathbf{n-1}|E_{j}\rangle=\delta_{n}\langle\mathbf{n+1}|E_{j}\rangle.\label{eq:eqn}\end{equation}
For $n=1$ this reads\[
\left(E_{j}-D_{1}\right)\langle\mathbf{1}|E_{j}\rangle=\delta_{1}\langle\mathbf{2}|E_{j}\rangle.\]
Since the l.h.s. is known for all $j$, the expansion of $|\mathbf{2}\rangle$
in the basis $|E_{j}\rangle$ is known up to the unknown constant
$\delta_{1}$ . Through normalization of $|\mathbf{2}\rangle$ we
then obtain $\left|\delta_{1}\right|$ and since the sign of $\delta_{1}$
is known by assumption, we obtain $\delta_{1}.$ Also, we now know
$\langle\mathbf{2}|E_{j}\rangle$ for all $j.$ The next equation
is obtained by setting $n=2$ in (\ref{eq:eqn}) as \[
\left(E_{j}-D_{2}\right)\langle\mathbf{2}|E_{j}\rangle-\delta_{1}\langle\mathbf{1}|E_{j}\rangle=\delta_{2}\langle\mathbf{3}|E_{j}\rangle.\]
The only unknown on the l.h.s. is $D_{2}$ which is obtained by Eq.~(\ref{eq:Dn}).
Using again the normalization we obtain $\delta_{2}$ and $\langle\mathbf{3}|E_{j}\rangle.$
We could continue this procedure, but the normalization would not
provide us with the signs of the $\delta_{n}.$ We know more: from
$D_{1}-D_{2}=-4\Delta\delta_{2}$ we obtain $\Delta$ and from $D_{1}$
we can then get $\sum_{m=1}^{N-1}\delta_{m}.$ Then, Eq.~(\ref{eq:Dn})
gives us $D_{3}$ and therefore $\delta_{3}$ \emph{including its
sign}. This method is then easily generalized by using Eq. (\ref{eq:eqn})
for $n=4$ to obtain $\langle\mathbf{3}|E_{j}\rangle,$ and Eq.~(\ref{eq:Dn})
for $\delta_{4},$ and so on.~$\blacksquare$ 

Now we describe how the requirements of the lemma can be measured
by controlling the first spin only. First, we initialize the system
in $\frac{1}{\sqrt{2}}\left(|\mathbf{0}\rangle+|\mathbf{1}\rangle\right)$.
We remark that this can be done by acting on the first spin only,
cf. \cite{Burgarth2007c}). We then perform quantum state tomography
\cite{NIELSEN} on the same spin at a later time. The evolved state
is \[
\frac{1}{\sqrt{2}}U(t)\left(|\mathbf{0}\rangle+|\mathbf{1}\rangle\right)=\frac{1}{\sqrt{2}}\left(\exp\left\{ -iGt\right\} |\mathbf{0}\rangle+\sum_{n=1}^{N}f_{n1}(t)|\mathbf{n}\rangle\right)\]
with $f_{n1}$ given by $\langle\mathbf{n}|U(t)|\mathbf{1}\rangle.$
The reduced density matrix of the first spin of the chain is given
by \[
\rho_{1}=\left(\begin{array}{cc}
2-\left|f_{11}\right|^{2} & \exp\{iGt\}f_{11}^{*}\\
\exp\{-iGt\}f_{11} & \left|f_{11}\right|^{2}\end{array}\right)/2.\]
 Thus, by performing quantum state tomography on spin $1$ we can
sample the following matrix element of the time evolution operator,\begin{equation}
\exp\{-Gt\}\langle\mathbf{1}|U(t)|\mathbf{1}\rangle=\sum_{j}\left|\langle\mathbf{1}\right|E_{j}\rangle|^{2}\exp\left\{ -i\left(E_{j}+G\right)t\right\} .\label{eq:totrans}\end{equation}
 Since the eigenvectors of $H$ are determined only up to an irrelevant
phase, we can assume, without loss of generality, that \[
\langle\mathbf{1}|E_{j}\rangle>0\,(j=1,\ldots,N).\]
 Hence through Fourier transforming Eq. (\ref{eq:totrans}), the coefficients
$\langle\mathbf{1}|E_{j}\rangle$ are known (as long as the spectrum
is non-degenerate). The eigenvalues $E_{j}$ are known up to the constant
$G.$ As usual, this global shift of eigenvalues does not change the
physics -- Eq. (\ref{eq:eqn}) shows that $G$ cancels out when determining
the coupling strengths. %
\begin{figure}
\includegraphics[width=1\columnwidth]{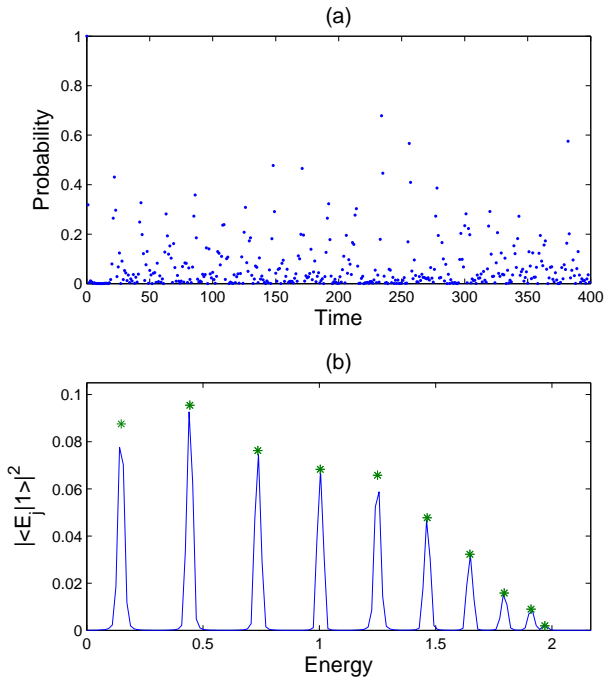}

\caption{(Color Online) \emph{\label{fig:example}}Simulated measurement data and its Fourier
transform. The green stars show the position of the exact eigenfrequencies.
The coupling strengths have been chosen randomly in the interval $1\pm0.05$
for a chain of $N=20.$ The simulated measurement points allowed the
estimation with a standard deviation of $0.01$ with respect to the
real couplings. The Fourier transform was computed using standard
FFT algorithms and a Hann window. Only the $10$ peaks with positive
$E_{j}$ are shown, the others are symmetric around the $y$-axis.}

\end{figure}

\section{Efficiency\label{sec:Efficiency}}

The efficiency of the coupling estimation can be studied using standard
properties of the Fourier transform (see \cite{Bracewell} for an
introduction). The functions $\langle\mathbf{1}|U(t)|\mathbf{1}\rangle$
is sampled for each time $t$ by state preparation, system evolution,
and quantum state tomography. Therefore an important cost parameter
is the total number of measured points, being proportional to the
sampling frequency. The minimal sampling frequency is given by the
celebrated Nyquist\textendash{}Shannon sampling theorem as $2f_{\mbox{min}}=E_{\mbox{max}}$,
where $E_{\mbox{max}}$ is the maximal eigenvalue of $H$ in the first
excitation sector. Due to decoherence and dissipation, the other important
parameter is the total time interval $T$ over which the functions
need to be sampled to obtain a good resolution. This is given by the
(classical) uncertainty principle that states that the frequency resolution
is proportional to $1/T.$ Hence the minimal time interval over which
we should sample scales as $T_{\mbox{min}}=1/\left(\Delta E\right)_{\mbox{min}},$
where $\left(\Delta E\right)_{\mbox{min}}$ is the minimal energy
distance of the eigenvalues of the Hamiltonian. Another important parameter
is the height of the peaks in the Fourier transform given by $\left|\langle\mathbf{1}|E_{j}\rangle\right|^{2}.$
These should be high enough to resolve, which means that all energy
eigenstates need to be well delocalized. If there is too much disorder,
localization will take place (see for example \cite{Burrel2007}),
and couplings far away from the controlled region can no longer be
probed (in turn, this suggests a way of obtaining information on localization
lengths indirectly - see conclusion). When localization is negligible,
the numerical algorithm to obtain the coupling strengths from the
Fourier transform is very stable \cite{Gladwell2004}. The reason is that
the couplings are obtained from a linear system of equations, so
errors in the quantum state tomography or effects of noise degrade the 
estimation linearly. In our numerical
analysis we found a good agreement with the real couplings for systems
with small randomness (Fig. \ref{fig:example}). Let us also look
at the scaling of the problem with the number of spins. Typically
the dispersion relationship in 1D systems of length $N$ is $\cos\frac{\pi k}{N},$
which means that the minimal energy difference scales as $\approx\frac{1}{N^{2}}$
and the total time interval should be chosen as $N^{2}.$ This agrees
well with our numerical results (tested up to $N=100$). For
each sampling point a quantum state tomography of a signal of an average height 
of $\approx 1/N$ needs to be performed. Since the error of tomography scales
inverse proportionally to the square root of the number of measurements, roughly $N^2$ measurements
are required for each tomography.

\section{Conclusion and Outlook}

In conclusion, we have found that a vast class of spin chain Hamiltonians
can be estimated by some restricted operation at the chain end. There
are obviously many variants of the inverse problem that we have introduced
above. For example, it is straightforward to see that the anisotropy
parameter $\Delta$ can be estimated even if it is site dependent
(i.e., $\Delta_{n}$) if one assumes that the signs of \emph{all} $\delta_{n}$
are known. We have emphasized here only one example as we think it
is the most relevant one for the applications \cite{Burgarth2008,Haselgrove}
and stands as a paradigm for the setup we have introduced: quantum
system identification of a \emph{black box }by restricted access.
There are many interesting more general and fundamental questions
that our study suggests. For instance, is it possible to count the
number of qubits by restricted access \cite{counting}? What can be said
about more complicated networks and higher dimensional lattices? If the black
box contains some dissipation and decoherence, what can we learn about
it through restricted access? What can be learned about the localization
length?

\begin{acknowledgments}
DB acknowledges the QIP-IRC and Wolfson College, Oxford. KM is grateful
for the support by the Incentive Research Grant of RIKEN. This work
is supported in part by the NSA, LPS, ARO, and the NSF grant No. EIA-0130383. 
\end{acknowledgments}

\end{document}